\documentclass[12pt]{article}
\usepackage{graphics, color}
\usepackage{amssymb}
\usepackage{amsmath}
\begin{document}

\baselineskip=15.5pt
\pagestyle{plain}
\setcounter{page}{1}


\begin{titlepage}
\bigskip
\rightline{hep-th/0508064}
\bigskip\bigskip\bigskip
\centerline{\Large \bf Integrable Open Spin Chains and the Doubling Trick}
\medskip
\centerline{\Large \bf in $\mathcal{N} = 2$ SYM with Fundamental Matter}
\bigskip\bigskip\bigskip


\centerline{\large Theodore G. Erler$^a$ and Nelia Mann$^b$}
\bigskip
\centerline{\em $^a$ Department of Physics, University of California} \centerline{\em Santa Barbara, CA 93106} \centerline{\em terler@physics.ucsb.edu}
\smallskip
\centerline{\em $^b$ Department of Physics, University of California} \centerline{\em Santa Barbara, CA 93106} \centerline{\em nelia@physics.ucsb.edu}
\bigskip
\bigskip
\bigskip\bigskip


\begin{abstract}
We demonstrate that the one-loop anomalous dimension matrix in $\mathcal{N} = 2$ SYM with a single chiral hypermultiplet of fundamental matter, which is dual to AdS$_{5} \times $S$^{5}$ with a D7-brane filling AdS$_{5}$ and wrapped around an S$^{3}$ in the S$^{5}$, is an integrable open spin chain Hamiltonian.  We also use the doubling trick to relate these open spin chains to closed spin chains in pure $\mathcal{N} = 4$ SYM.  By using the AdS/CFT correspondence, we find a relation between the corresponding open and closed strings that differs from a simple doubling trick by terms that vanish in the semiclassical limit.  We also demonstrate that in some cases the closed string is simpler and easier to study than the corresponding open string, and we speculate on the nature of corrections due to the presence of D-branes that this implies.
\medskip
\noindent
\end{abstract}
\end{titlepage}

\section{Introduction}

One of the most interesting subjects in string theory has been the conjectured equivalence between string theory on $AdS_{5} \times\mathrm{S}^{5}$ and an $\mathcal{N} = 4$ $SU(N)$ super Yang-Mills theory \cite{Maldacena}-\cite{MAGOO}. Though many aspects of this proposed correspondence remain a mystery, we can make much progress by restricting attention to certain sectors, such as operators with large R-charge in the CFT, which on the gravity side of the duality correspond to strings in a plane wave background \cite{BMN}.  In these sectors it is particularly easy to see how a closed string arises as a chain of ``bits,'' dual to a gauge invariant single-trace operator.

We have also come to understand that there is a rich structure of integrability on both sides of the correspondence.  On the CFT side, Minahan and Zarembo showed that the 1-loop anomalous dimension matrix for single trace scalar operators can be diagonalized using Bethe ansatz techniques \cite{MZ}; this work was further extended to the full CFT and higher loops in subsequent papers \cite{BS}-\cite{Roiban}. The resulting spectrum has been successfully matched to energies of corresponding semiclassical string states \cite{Beis}.  Meanwhile, on the $AdS$ side, it has been shown\cite{BPR} that the string sigma model on $AdS_{5}\times\mathrm{S}^{5}$ possesses an infinite number of Yangian symmetries, suggesting the possibility that the theory might in fact be exactly solvable.  Work to demonstrate the integrability of the AdS string has continued in \cite{Vallilo:2003nx}-\cite{Alday:2003zb}, and relations between the integrable structures on both sides have been proposed \cite{DNW}.  More recent work on integrability and AdS/CFT includes \cite{Arutyunov:2003uj}-\cite{quant/finitesize2}.  

Integrability has also been studied in deformations that add fundamental matter into the CFT, which correspond to open strings living on a D-brane in $AdS$ \cite{CWW}-\cite{giantgrav}. These deformations have been further related to semiclassical open spinning strings \cite{openstringsdCFT}-\cite{openstrings}.   In this note we consider another nice example of integrability in the open string sector: we consider a probe D7-brane filling all of the AdS, which adds a single $\mathcal{N} = 2$ chiral hypermultiplet of fundamental matter to the CFT \cite{Karch:2002sh}-\cite{Bab}. We compute the one loop anomalous dimension matrix and compute its spectrum using Bethe ansatz techniques. This system is particularly natural since the fundamental matter is free to propagate in all four dimensions, thus we avoid  some subtleties of the defect in ref. \cite{ME} while at the same time working with a system that more closely resembles QCD. The D7-brane system is also further away from pure $\mathcal{N} = 4$ SYM in the sense that the conformal symmetry is broken except in the strict large $N$ limit, where the D7 brane acts as a probe on the AdS.  Interestingly, one loop integrability is preserved.  Furthermore, we can use this system to study more carefully techniques relating open spin chains to closed spin chains \cite{OSC/OSS} and how these techniques relate to a traditional string ``doubling trick''.  The analysis of this system has similarities to the work with the defect CFT of ref. \cite{ME}; both allow for operators corresponding to open spin chains whose boundaries break the $SO(6)$ R-symmetry down to a subgroup.

 We begin in section 1 with a field theoretic determination of the one-loop anomalous dimension matrix, and demonstrate that it vanishes when acting on our chiral primary operators.  In Section 2 we then determine the integrable spin chain with the same symmetries as our operators, and demonstrate that our anomalous dimension matrix is one of the complete set of commuting operators for this system.  Having demonstrated that our anomalous dimension matrix can be diagonalized using a a Bethe ansatz, we then find the Bethe ansatz in section 3, by studying spin chains with single impurities.  Our analysis in this section confirms many of the features discovered in the previous paper using the dCFT. ~\cite{ME} 
 
In section 4, we move forward with both the open spin chains in the $\mathcal{N} = 2$ system studied in the previous sections, and the open spin chains in the dCFT studied in ~\cite{ME}.  In ~\cite{OSC/OSS}, it was shown that Bethe ansaetze for open spin chains can be directly related to closed spin chain Bethe ansaetze by a version of the ``doubling trick.''  We apply this method to our open spin chains, and using the AdS/CFT correspondence, we then find the related open and closed strings.  Our relationships between open strings and closed strings have two interesting features that go beyond a simple doubling trick.  We find that the relation differs from the expected doubling trick by terms of order $\mathcal{O}1/J$, where $J$ is the total angular momentum of the string, or even by terms of order $\mathcal{O}(1)$.  This first type of correction is obscured in semiclassical analyses, and reflects the effect of boundary conditions on the energy of an open string.  The second reflects the care that needs to be taken in correctly identifying the appropriate closed string to relate to the open string.  In this case we will find open strings that do not satisfy $E - J = \mathcal{O}(\lambda/J^2)$, but are related by ``doubling'' to closed strings that do.   Conclusions and open questions are discussed in section 5, and an appendix includes field theory conventions. 
 
\section{The Anomalous Dimension Matrix}

\subsection{The $\mathcal{N} = 2$ Action}

The field theory we study here is a variation of $\mathcal{N} = 4, SU(N)$ SYM in which we add one $\mathcal{N} = 2$ hypermultiplet of fundamental matter.  This breaks the supersymmetry down to $\mathcal{N} = 2$, and breaks the R-symmetry from $SO(6) = SU(4)$ to $SO(4) \times SO(2) = SU(2)_{L} \times SU(2)_{R} \times U(1)$.  In addition, the conformal symmetry is anomalous at finite $N$, though it is restored in the strict large $N$ limit where we will be studying the theory.  The content of this theory is then the content of the $\mathcal{N} = 4$ theory (6 real scalars, 4 Weyl fermions, and one gauge boson, all in the adjoint of the gauge group), and a hypermultiplet of matter (an $SU(2)$ doublet of complex scalars and two Weyl fermions, in the fundamental of the gauge group) \cite{mesons}.
The pure $\mathcal{N} = 4$ action
\begin{equation}
S_{\mathcal{N} = 4} = \int d^4x Tr\left\{ -\frac{1}{4}F^{\mu\nu}F_{\mu\nu} - i\psi_{A}\sigma^{\mu}D_{\mu}\bar{\psi}^{A} - \frac{1}{2}D^{\mu}\phi^{i}D_{\mu}\phi^{i} \right.
\end{equation}
\begin{displaymath}
\left.+ g\phi\psi_{A}C_{i}^{AB}\psi_{B}  + g\phi_{i}\bar{\psi}^{A}\bar{C}^{i}_{AB}\bar{\psi}^{B} + \frac{g^2}{4}[\phi_{i}, \phi_{j}]^2\right\},
\end{displaymath}
where the $C_{i}^{AB}$ are Clebsch-Gordon matrices translating between the $4$ and $6$ representations of the R-symmetry group $SU(4)$, is supplemented by the additional terms
\begin{equation}
S_{fund} = \int d^4x \Bigg\{ -(D^{\mu}Q^{a})^{\dagger}D_{\mu}Q^{a} - i\bar{\chi}\bar{\sigma}^{\mu}D_{\mu}\chi - i\pi\sigma^{\mu}D_{\mu}\bar{\pi}\Bigg.
\end{equation}

\begin{displaymath}
\left. -ig\sqrt{2}\pi Z \chi + ig\sqrt{2}\bar{\chi}\bar{Z}\bar{\pi} - \frac{g^2}{2}(\bar{Q}_{a}Q^{b})(\bar{Q}_{b}Q^{a}) - g^2\epsilon_{ab}\epsilon^{cd}(\bar{Q}_{c}Q^{b})(\bar{Q}_{d}Q^{a})\right.
\end{displaymath}

\begin{displaymath}
\left. + ig\sqrt{2}\bar{Q}_{a}\bar{\Lambda}^{a}\bar{\pi} - ig\sqrt{2}\pi\Lambda_{a}Q^{a} + ig\sqrt{2}\bar{Q}_{a}\epsilon^{ab}\Lambda_{b}\chi - ig\sqrt{2}\bar{\chi}\bar{\Lambda}^{a}\epsilon_{ab}Q^{b}\right.
\end{displaymath}

\begin{displaymath}
\Bigg. - g^2\bar{Q}_{a}\phi_{i}\phi_{i}Q^{a} + g^2\bar{Q}_{b}\phi^{I}\bar{W}_{I  a\bar{c}}W_{J}^{\bar{c} b}\phi^{J}Q^{a} \Bigg\}.
\end{displaymath}

Here, $Q$ is an $SU(2)_{R}$ doublet of complex fundamental scalars, $\pi$ and $\chi$ are fundamental fermions, $Z$ is a complex combination of two real adjoint scalars (charged under the $U(1)$), $\Lambda$ is the doublet of adjoint fermions that transform under $SU(2)_{R}$, and $W_{I}^{\bar{a}b}$ are Clebsch-Gordon matrices that translate between the (1/2,1/2) of $SU(2)_{L} \times SU(2)_{R}$ and the $4$ of $SO(4)$.  In the appendix there are tables summarizing the charges of all fields, as well as full explanations of our field theory conventions.

\subsection{Feynman Diagram Calculations}

We would like to find the anomalous dimension matrix for scalar operators of the type 
\begin{equation}
\mathcal{O} = \bar{Q}_{a_1}\phi^{i_1}\cdots\phi^{i_{L}}Q^{a_2}.
\end{equation}
To one loop, and in the strict large $N$ limit, this limits us to ``nearest neighbor'' interactions.  Interactions between two $\phi^{i}$ fields are identical to those encountered in one-loop anomalous dimension matrices in $\mathcal{N} = 4$ SYM because terms involving fundamental matter will always be suppressed by orders of $1/N$.  Boundary interactions between one fundamental and one adjoint scalar should, at this order, always yield one fundamental and one adjoint scalar.  Thus, we expect these operators only to mix among themselves.  The matrix should act on a Hilbert space
\begin{equation}
\mathbb{C}^{2} \times \underbrace{\mathbb{R}^{6} \times \cdots \times \mathbb{R}^{6}}_{L \ \mbox{copies}} \times \mathbb{C}^{2}
\end{equation}
and consist of ``nearest neighbor interactions''.  The interior terms should be exactly the usual $\mathcal{N} = 4$ SYM 1-loop anomalous dimension terms, and these should be supplemented by operators giving interactions between the boundary $\mathbb{C}^{2}$s and the $\mathbb{R}^{6}$s next to them.

The anomalous dimension matrix is calculated using the usual Feynman diagram approach \cite{MZ}.  We find that if the operator is renormalized to be
\begin{equation}
\mathcal{O}^{A}_{ren} = Z^{A}_{B}\mathcal{O}^{B}
\end{equation}
so that correlation functions with this operator are finite, then the anomalous dimension matrix is
\begin{equation}
\Gamma = \frac{dZ}{d\ln \Lambda}\cdot Z^{-1},
\end{equation}
where $\Lambda$ is the UV cutoff.  We thus want to calculate the correlation function
\begin{equation}
\langle Q^{a_1}\phi^{I_1}\cdots \phi^{I_L}\bar{Q}_{a_2} \mathcal{O}\rangle
\end{equation}
at one loop, which will involve corrections to each of the fields' propagators as well as nearest-neighbor type terms between two adjacent scalar fields.

In addition to the normal types of fields we have in $\mathcal{N} = 4$,

\resizebox{6in}{!}{\includegraphics{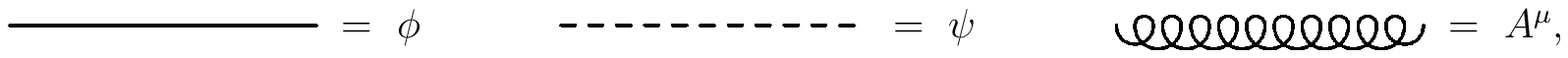}}

there are three fundamental field types

\resizebox{6in}{!}{\includegraphics{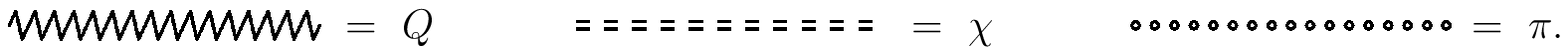}}

We start out with  the usual $\mathcal{N} = 4$ diagrams.  First we have  the one-loop correction to the adjoint scalar propagator.

\resizebox{6in}{!}{\includegraphics{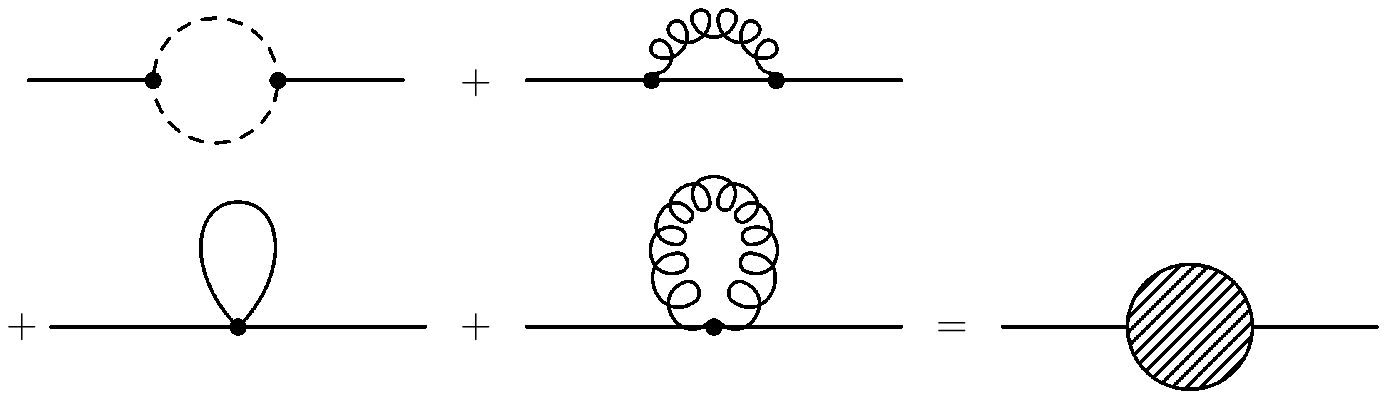}}

Note that the two diagrams involving quartic interactions don't directly contribute to the anomalous dimension; they help cancel quadratic divergences in the other diagrams whose presence would indicate a dynamically generated mass term.  If we considered contributions away from the strict large-N limit, we would also have diagrams here involving the fundamental matter; these would not have canceling quadratic divergences, and thus would indicate an anomaly in the conformal symmetry.

We also have the exchange of a gluon, and a four-scalar interaction among four adjoint scalars

\resizebox{6in}{!}{\includegraphics{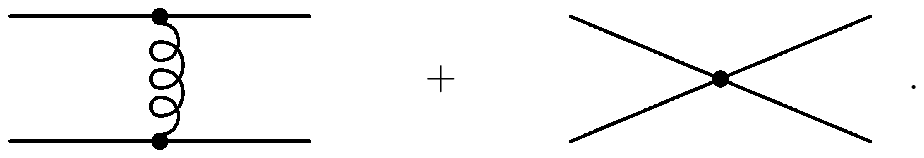}}

These were calculated in \cite{MZ}.  

From the two ends of the operator we have the one-loop corrections to the fundamental scalar

\resizebox{6in}{!}{\includegraphics{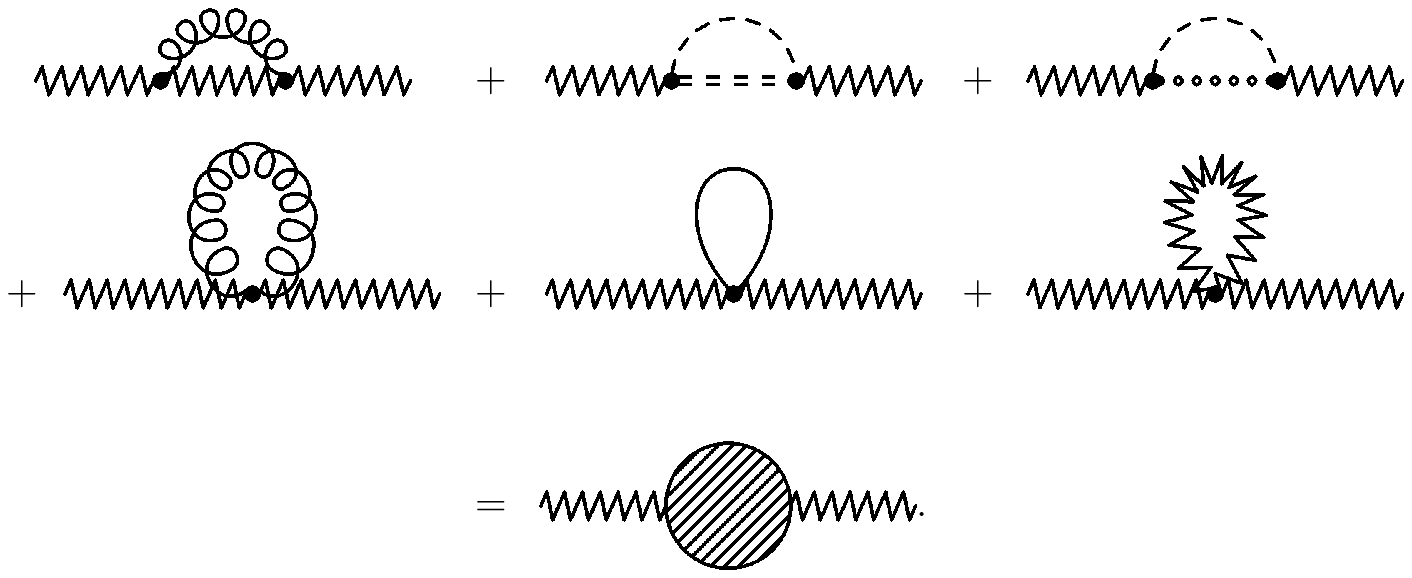}}

(Again, the quartic terms don't directly contribute to the anomalous dimension)
as well as a gluon exchange between a fundamental and an adjoint scalar, and a four-scalar interaction between two fundamentals and two adjoints.

\resizebox{6in}{!}{\includegraphics{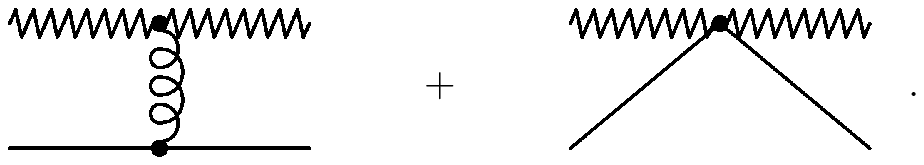}}

After calculating these Feynman diagrams, we find that the anomalous dimension matrix is
\begin{equation}\label{eq:gamma}
\Gamma = \frac{g^2N}{16\pi^2}\left[R_{1} - 2I + \bar{R}_{L} - 2I + \sum_{n = 1}^{L-1} (2P_{n,n+1} - 2I_{n,n+1} - K_{n,n+1})\right]
\end{equation}
where the operators acting on the interior of the spin chain are defined, as usual, to be
\begin{eqnarray}\label{eq:PIK}
(P_{n,n+1})^{b_1, j_1, ..., j_L, a_2}_{a_1, i_1, ..., i_L, b_2} & = & \delta^{b_1}_{a_1}\delta^{j_1}_{i_1} \cdots \delta^{j_{n-1}}_{i_{1}} (\delta^{j_n}_{i_{n+1}}\delta^{j_{n+1}}_{i_{n}}) \delta^{j_{n+2}}_{i_{n+2}}\cdots \delta^{j_{L}}_{i_{L}}\delta^{a_2}_{b_2} \nonumber \\
(I_{n,n+1})^{b_1, j_1, ..., j_L, a_2}_{a_1, i_1, ..., i_L, b_2} & = & \delta^{b_1}_{a_1}\delta^{j_1}_{i_1} \cdots \delta^{j_{n-1}}_{i_{1}} (\delta^{j_n}_{i_{n}}\delta^{j_{n+1}}_{i_{n+1}}) \delta^{j_{n+2}}_{i_{n+2}}\cdots \delta^{j_{L}}_{i_{L}}\delta^{a_2}_{b_2} \nonumber \\
(K_{n,n+1})^{b_1, j_1, ..., j_L, a_2}_{a_1, i_1, ..., i_L, b_2} & = & \delta^{b_1}_{a_1}\delta^{j_1}_{i_1} \cdots \delta^{j_{n-1}}_{i_{1}} (\delta^{j_n, j_{n+1}}\delta_{i_n, i_{n+1}}) \delta^{j_{n+2}}_{i_{n+2}}\cdots \delta^{j_{L}}_{i_{L}}\delta^{a_2}_{b_2}.
\end{eqnarray}
The operators that act on the ends of the spin chain are defined as
\begin{eqnarray}\label{eq:Rdef}
(R_{1})^{b_1, j_1, ..., j_L, a_2}_{a_1, i_1, ..., i_L, b_2} & = & (\delta_{i_1, I}\delta^{j_1, J}\bar{W}_{J a_1\bar{c}}W_{I}^{\bar{c} b_1}) \delta^{j_2}_{i_2} \cdots \delta^{j_{L}}_{i_{L}}\delta^{a_2}_{b_2} \nonumber \\
(\bar{R}_{L})^{b_1, j_1, ..., j_L, a_2}_{a_1, i_1, ..., i_L, b_2} & = & \delta^{j_1}_{i_1} \cdots \delta^{j_{L-1}}_{i_{L-1}}(\delta_{i_L, I}\delta^{j_L, J}\bar{W}_{I b_2 \bar{c}}W_{J}^{\bar{c} a_2}).
\end{eqnarray}

Note that the nearest neighbor terms from the interior of the operator are identical to those calculated in \cite{MZ}.

\subsection{The Chiral Primary Operators}

As a check on our anomalous dimension matrix, we find a chiral primary operator of this type, and show that it has vanishing anomalous dimension (at least to one-loop).  We know that in $\mathcal{N} = 4$ SYM, operators of the type $Tr (\phi_{i} + i\phi_{j})^{L}$ for $i \ne j$ are CPOs- they have maximal charge under one generator of $SO(6)$.  Equivalently, we want to look for operators of our type with maximal charge.  Of course, since the fields $Q^{a}$ and $\bar{Q}_{a}$ are uncharged under the $SO(2) = U(1)$, we need to restrict ourselves to adjoint fields charged under the $SO(4)$.  For example, suppose we look for an operator with maximal charge under rotations in the $1-2$ plane, so that $\phi_{1} + i\phi_{2}$ obtains a positive charge $+1$.  The SO(6) matrix for which this vector has eigenvalue $+1$ is
\begin{equation}
T = -i\left(\begin{array}{cccccc} 0 & 1 & 0 & 0 & 0 & 0 \\ -1 & 0 & 0 & 0 & 0 & 0 \\ 0 & 0 & 0 & 0 & 0 & 0 \\ 0 & 0 & 0 & 0 & 0 & 0 \\ 0 & 0 & 0 & 0 & 0 & 0 \\ 0 & 0 & 0 & 0 & 0 & 0 \end{array}\right).
\end{equation}
The corresponding generator for the $4$ of $SU(4)$ can be found by contracting the matrix $T^{i}_{j}$ with the Clebsch-Gordan matrices $C_{i}^{AB}$:
\begin{equation}
\frac{1}{4}T^{i}_{j}C_{i}\bar{C}^{j} = \frac{1}{2}\left(\begin{array}{cccc} 1 & 0 & 0 & 0 \\ 0 & -1 & 0 & 0 \\ 0 & 0 & -1 & 0 \\ 0 & 0 & 0 & 1 \end{array}\right).
\end{equation}
Now, this matrix shows that $\Lambda_{1}$ would have charge $-1/2$ and $\Lambda_{2}$ would have charge $1/2$, while $\Theta_{1}$ would have charge $1/2$ and $\Theta_{2}$ would have charge $-1/2$.  Since $\bar{Q}$ is in the same representation of the same $SU(2)$ as $\Lambda$, this tells us that $\bar{Q}_{2}$ has charge $1/2$ under this rotation.  If we transform the $SU(2)$ generator $\sigma_{3}$ into the generator acting on an anti-fundamental object, we find that $Q^{1}$ also has charge $1/2$.  Thus, our CPO with maximal charge under this rotation should be
\begin{equation}
\bar{Q}_{2}(\phi_{1} + i\phi_{2})^{L}Q^{1},
\end{equation}
and of course there are infinitely more such CPOs.  Now, consider our anomalous dimension matrix acting on the operator $\mathcal{O} = \bar{Q}_{2}(\phi_{1} + i\phi_{2})^{L}Q^{1}$.  We know that each term acting on the interior of the operator $2P_{n,n+1} - 2I_{n,n+1} - K_{n,n+1}$ vanishes independently, just as it does for the single trace operator of $\mathcal{N} = 4$ SYM.  The terms $R_{1} - 2I$ and $\bar{R}_{L} - 2I$ also each vanish independently:
\begin{eqnarray}
\lefteqn{(R_{1} - 2I)\mathcal{O}} \nonumber \\
 & = & \bar{W}_{J 2\bar{c}}(W_{1} + iW_{2})^{\bar{c}b_{1}}\bar{Q}_{b_{1}}\phi_{J}(\phi_{1} + i\phi_{2})^{L - 1}Q^{1} - 2\bar{Q}_{2}(\phi_{1} + i\phi_{2})^{L}Q^{1} \nonumber \\
& = & -2\bar{W}_{J 21}\bar{Q}_{2}\phi_{J}(\phi_{1} + i\phi_{2})^{L - 1}Q^{1} - 2\bar{Q}_{2}(\phi_{1} + i\phi_{2})^{L}Q^{1} \nonumber \\
& = & 2\bar{Q}_{2}(\phi_{1} + i\phi_{2})^{L}Q^{1} - 2\bar{Q}_{2}(\phi_{1} + i\phi_{2})^{L}Q^{1} \nonumber \\
& = & 0
\end{eqnarray}
and
\begin{eqnarray}
\lefteqn{(\bar{R}_{L} - 2I)\mathcal{O}} \nonumber \\
& = & (\bar{W}_{1} + i\bar{W}_{2})_{b_{2}\bar{c}}W_{J}^{\bar{c}1}\bar{Q}_{2}(\phi_{1} + i\phi_{2})^{L - 1}\phi_{J}Q^{b_{2}} - 2\bar{Q}_{2}(\phi_{1} + i\phi_{2})^{L}Q^{1} \nonumber \\
& = & -2W_{J}^{21}\bar{Q}_{2}(\phi_{1} + i\phi_{2})^{L - 1}\phi_{J}Q^{1} - 2\bar{Q}_{2}(\phi_{1} + i\phi_{2})^{L}Q^{1} \nonumber \\
& = & 2\bar{Q}_{2}(\phi_{1} + i\phi_{2})^{L}Q^{1} - 2\bar{Q}_{2}(\phi_{1} + i\phi_{2})^{L}Q^{1} \nonumber \\
& = & 0.
\end{eqnarray}
Thus, our anomalous dimension matrix does vanish when acting on our CPOs, just as it should.  From now on we will make the definitions
\begin{equation}
X = \phi_{1} + i\phi_{2}, \ \ \ \ \ Y = \phi_{3} + i\phi_{4}.
\end{equation}
We will use the operator $\bar{Q}_{2}X^{L}Q^{1}$ as a reference operator, and study impurities of the types $Y$, $\bar{Y}$, and $Z = \phi_{5} + i\phi_{6}$ in it.  Note that if we write the charges associated with $X$, $Y$, $Z$ for an operator as $(J_1, J_2, J_3)$, this reference operator has charges $(L+1, 1, 0)$, because the fundamental matter is charged under two $SO(2)$ subgroups.

\section{The Boundary Yang-Baxter Equation}

We would now like to determine whether or not the anomalous dimension matrix calculated above is one of an infinite number of commuting operators in an integrable system.  These calculations follow closely the procedure used in \cite{ME}, but are spelled out here for convenience.

\begin{figure}
\resizebox{!}{1in}{\includegraphics{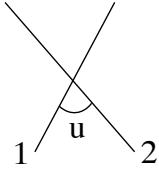}}
\caption{\label{R-matrix} The $\mathcal{R}_{12}(u)$ matrix (equation (\ref{eq:R-matrix})), represented graphically.}
\end{figure}

\begin{figure}
\resizebox{!}{1.5in}{\includegraphics{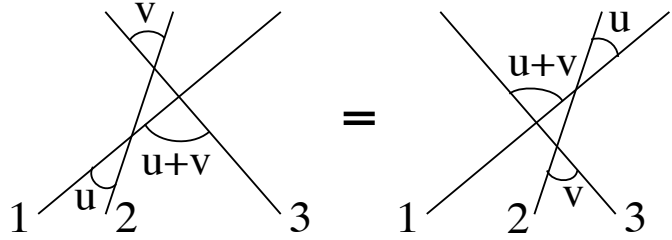}}
\caption{\label{YBeq} The Yang-Baxter equation (\ref{eq:YB}), graphically.}
\end{figure}

\begin{figure}
\resizebox{!}{1.5in}{\includegraphics{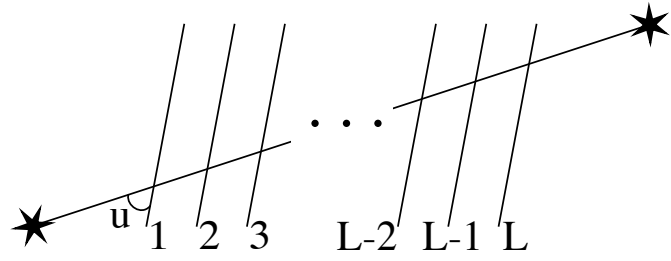}}
\caption{\label{transfer} The transfer matrix $t(u)$ defined in equation (\ref{eq:transfer}), graphically.  The stars are meant to be identified with each other to represent the trace.}
\end{figure}

\begin{figure}
\resizebox{6in}{!}{\includegraphics{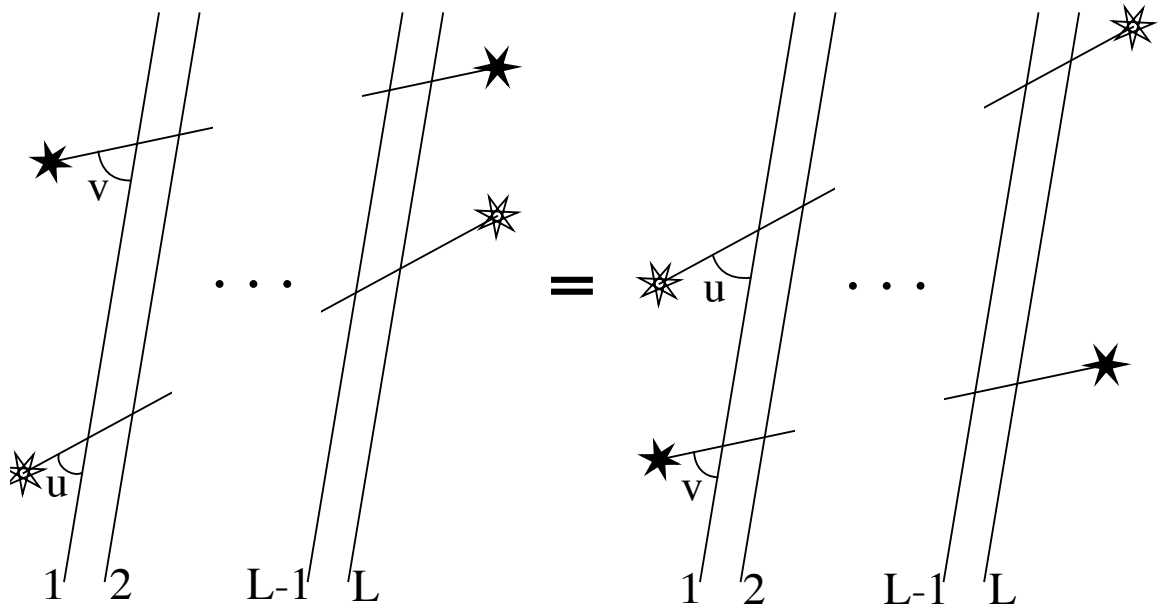}}
\caption{\label{commute} The commutation of two transfer matrices, graphically.  Equation (\ref{eq:commute}).}
\end{figure}

For closed spin chains, Minahan and Zarembo showed in \cite{MZ} that operators composed of closed chains of adjoint scalar fields do have this structure.  They did this by identifying an $SO(6)$ invariant $\mathcal{R}$-matrix
\begin{equation}\label{eq:R-matrix}
\mathcal{R}_{12}(u) = \frac{1}{2}\left[u(u-2)I_{12} - (u-2)P_{12} + uK_{12}\right],
\end{equation}
where $1$ and $2$ label the two vector spaces (spin chain sites) that the operator acts on, and $I$, $P$, and $K$ are as defined in equation (\ref{eq:PIK}).  These matrices satisfy the Yang-Baxter equation
\begin{equation}\label{eq:YB}
\mathcal{R}_{12}(u)\mathcal{R}_{13}(u+v)\mathcal{R}_{23}(v) = \mathcal{R}_{23}(v)\mathcal{R}_{13}(u+v)\mathcal{R}_{12}(u).
\end{equation}
We can express this graphically.  In figure \ref{R-matrix} we show the graphical representation of the $\mathcal{R}_{12}(u)$ matrix, and in figure \ref{YBeq} we show the graphical representation of the Yang-Baxter equation.
We can then define the transfer matrix as the trace of the monodromy matrix,
\begin{equation}\label{eq:transfer}
t(u) = \mbox{Tr}_{a} T_{a}(u) \equiv \mbox{Tr}_{a} \mathcal{R}_{a1}(u)\mathcal{R}_{a2}(u)\cdots\mathcal{R}_{aL}(u),
\end{equation}
where $a$ labels an auxiliary vector space of the same type as the sites in the spin chain.  Thus, the monodromy matrix acts on the vector space $V_{a} \times V_{1} \times \cdots \times V_{L}$ for a spin chain of length $L$, and the transfer matrix acts on the Hilbert space of the spin chain.  If the $\mathcal{R}$-matrix satisfies the Yang-Baxter equation (\ref{eq:YB}), then the transfer matrices satisfy the relation
\begin{equation}\label{eq:commute}
[t(u), t(v)] = 0,
\end{equation}
and thus we can expand $t(u)$ in $u$ to get a complete set of mutually commuting operators and the spin chain is integrable.  If one of these operators is interpreted as a Hamiltonian, the others then represent more conserved charges.  We show the transfer matrix in figure \ref{transfer} and this commutation relation in figure \ref{commute}.

The single-trace one-loop anomalous dimension matrix can be written as a linear combination of the conserved charges, so it is integrable.

\begin{figure}
\resizebox{!}{1in}{\includegraphics{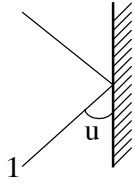}}
\caption{\label{K-matrix} The $\mathcal{K}^{-}(u)$-matrix, graphically.}
\end{figure}

\begin{figure}
\resizebox{!}{2in}{\includegraphics{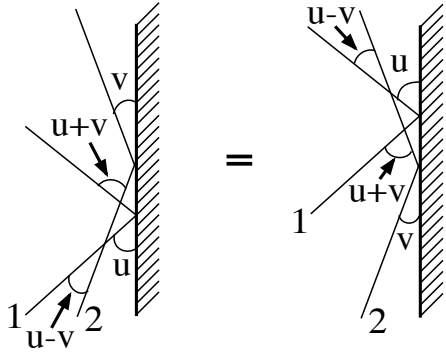}}
\caption{\label{BYB} A graphical representation of the BYB, equation (\ref{eq:BYB}).}
\end{figure}

\begin{figure}
\resizebox{!}{2in}{\includegraphics{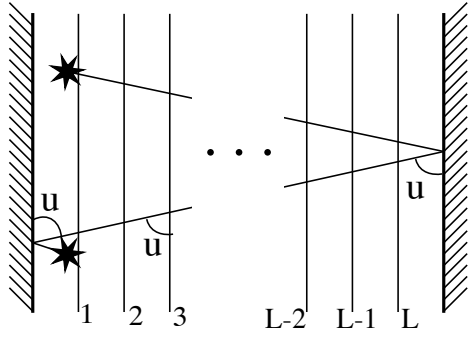}}
\caption{\label{trans2} The new transfer matrix $\hat{t}(u)$ equation (\ref{eq:trans2}), represented graphically.}
\end{figure}

\begin{figure}
\resizebox{6in}{!}{\includegraphics{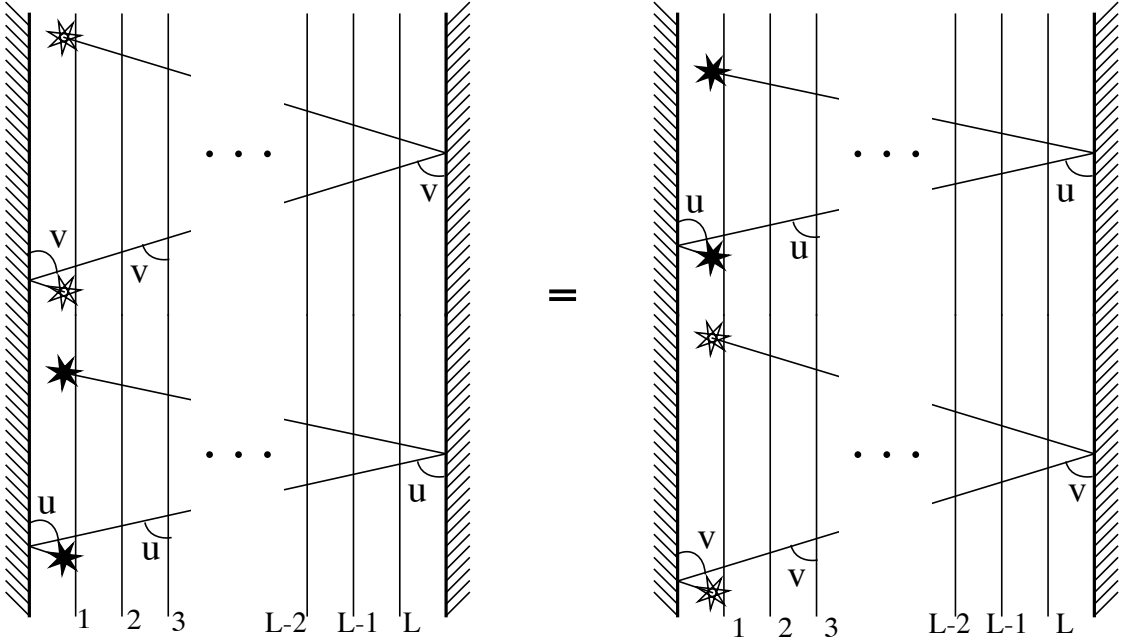}}
\caption{\label{commute2} The new transfer matrices commute with each other (equation (\ref{eq:commute2})).}
\end{figure}

Now, the analogous method for studying spin chains with boundaries was formulated in \cite{Skly}, and  was applied to the anomalous dimensions of operators in a superconformal field theory with fundamental matter in \cite{ME}.  The technique is to introduce matrices $\mathcal{K}^{\pm}_{a}(u)$ acting on either end of the spin chain in addition to the $\mathcal{R}_{12}(u)$ matrices.  In our case, these matrices will act on a vector space
\begin{equation}\label{eq:VtimesC}
\mathbb{R}^{6} \times \mathbb{C}^{2},
\end{equation}
though the action on the boundary $\mathbb{C}^{2}$ will be suppressed in the following equations.  The subscript $a$ in $\mathcal{K}^{\pm}_{a}(u)$ labels the copy of $\mathbb{R}^{6}$ that it acts on.
These satisfy the boundary Yang-Baxter equations (BYBs)
\begin{eqnarray}\label{eq:BYB}
\mathcal{R}(u-v)\mathcal{K}^{-}_{1}(u)\mathcal{R}_{12}(u+v)\mathcal{K}_{2}^{-}(v) & = & \mathcal{K}_{2}^{-}(v)\mathcal{R}_{12}(u+v)\mathcal{K}_{1}^{-}(u)\mathcal{R}_{12}(u-v) \\
\mathcal{R}_{12}(v-u)\mathcal{K}_{1}^{+t_{1}}(u)\mathcal{R}_{12}(-u-v-2i\gamma)\mathcal{K}_{2}^{+t_{2}}(v) & = & \mathcal{K}_{2}^{+t_{2}}\mathcal{R}_{12}(-u-v-2i\gamma)\mathcal{K}_{1}^{+t_{1}}(u)\mathcal{R}_{12}(v-u) \nonumber
\end{eqnarray}
where $t_{i}$ is a transpose on the $i$th vector space, and $\gamma$ is a parameter defined by the relation
\begin{equation}\label{inverse}
\mathcal{R}_{12}^{t_{1}}(u)\mathcal{R}_{12}^{t_{1}}(-u-2i\gamma) = \lambda(u)
\end{equation}
for $\lambda(u)$ a scalar function.  For our case we have $\gamma = 2i$.  In figure \ref{K-matrix} we show a graphical representation of the $\mathcal{K}^{-}$-matrix, and in figure \ref{BYB} we show the graphical BYB that it obeys  (the representations of the $\mathcal{K}^{+}$ are similar).
Now, we can define a new ``transfer'' matrix
\begin{equation}\label{eq:trans2}
\hat{t}(u) = \mbox{Tr}_{a} \mathcal{K}^{+}_{a}(u)T_{a}(u)\mathcal{K}^{-}_{a}(u)T^{-1}_{a}(u)
\end{equation}
that will act on the Hilbert space we are trying to study.  If both (\ref{eq:YB}) and (\ref{eq:BYB}) are satisfied, then these are the continuous set of mutually commuting operators needed for an integrable system:
\begin{equation}\label{eq:commute2}
[\hat{t}(u), \hat{t}(v)] = 0.
\end{equation}
The new transfer matrix is shown in figure \ref{trans2}, and the new commutation relation is shown in figure \ref{commute2}.

We need to find a matrix $\mathcal{K}^{-}_{a}(u)$ that satisfies equation (\ref{eq:BYB}) and preserves the symmetry group $SO(2) \times SO(4)$.  Again following the techniques of \cite{ME}, we make the ansatz
\begin{eqnarray}
(\mathcal{K}^{-})^{b J}_{a I}(u) & = & f(u)\delta^{b}_{a}\delta^{J}_{I} + g(u)\bar{W}^{J}_{a\bar{c}}W_{I}^{\bar{c}b} \nonumber \\
(\mathcal{K}^{-})^{b Y}_{a X}(u) & = & h(u)\delta^{b}_{a}\delta^{Y}_{X} \nonumber \\
(\mathcal{K}^{-})^{b Y}_{a I}(u) & = & 0 \nonumber \\
(\mathcal{K}^{-})^{b J}_{a X}(u) & = & 0,
\end{eqnarray}
remembering that $X,Y$ are indices in the $SO(2)$ and $I,J$ are indices in the $SO(4)$.  This is the most general ansatz we can make, using the group theory structure we have.  The BYB, written in index form, has four external $SO(6)$ indices $i_{1}, i_{2}, \ell_{1}, \ell_{2}$, and each index can be either in the $SO(2)$ or in the $SO(4)$.  It also has two external $SU(2)$ indices from acting on the boundary degrees of freedom. 
\begin{equation}\label{eq:BYBindices}
\mathcal{R}^{j_{1} j_{2}}_{i_{1} i_{2}}(u-v)(\mathcal{K}^{-})^{b k_{1}}_{a j_{1}}(u)\mathcal{R}^{\ell_{1} k_{2}}_{k_{1} j_{2}}(u + v)(\mathcal{K}^{-})^{c \ell_{2}}_{b k_{2}}(v) = (\mathcal{K}^{-})^{b j_{2}}_{a i_{2}}(v)\mathcal{R}^{j_{1} k_{2}}_{i_{1} j_{2}}(u + v)(\mathcal{K}^{-})^{c k_{1}}_{b j_{1}}(u)\mathcal{R}^{\ell_{1} \ell_{2}}_{k_{1} k_{2}}(u - v)
\end{equation}
Each case should be considered separately, and may give conditions that the functions $f(u)$, $g(u)$, and $h(u)$ need to satisfy.

For the eight cases where one index is in the $SO(2)$ and the other three are in the $SO(4)$, or vice-versa, both sides of the equation vanish and we get no conditions.  The case with all external $SO(6)$ indices in the $SO(2)$ is satisfied automatically, as is the case where $i_{1}$ and $\ell_{1}$ are in the $SO(2)$ and $i_{2}$ and $\ell_{2}$ in the $SO(4)$, or the opposite case where $i_{2}$ and $\ell_{2}$ are in the $SO(2)$ and $i_{1}$ and $\ell_{1}$ are in the $SO(4)$.

The cases where $i_{1}, \ell_{2} \in SO(2)$, $i_{2}, \ell_{1} \in SO(4)$ or the opposite $i_{2}, \ell_{1} \in SO(2)$, $i_{1}, \ell_{2} \in SO(4)$ each give us the equations
\begin{equation}
(u+v)(f(u)h(v) - f(v)h(u)) = (u-v)(4g(u)g(v) - f(v)f(u) - h(u)h(v)
\end{equation}
and
\begin{equation}
(u+v)(g(u)h(v) - g(v)h(u)) = (u-v)(g(v)f(u) + g(u)f(v)).
\end{equation}

The cases where $\ell_{1}, \ell_{2} \in SO(2)$, $i_{1}, i_{2} \in SO(4)$ or the opposite $i_{1}, i_{2} \in SO(2)$, $\ell_{1}, \ell_{2} \in SO(4)$ each give us the equations
\begin{eqnarray}
(u^2 - v^2)(u-v-2)(h(v)g(u) + h(u)g(v)) + (u+v)(u-v-2)(h(v)g(u) - h(u)g(v)) &&  \\
+ (u^2 - v^2)(u+v-2)(f(u)g(v) - f(v)g(u)) - (u-v)(u+v-2)(f(v)g(u) + f(u)g(v)) && \nonumber \\
+ 2(u+v)(u^2 - v^2)g(v)g(u) + 4(u^2 - v^2)f(u)g(v) + 2(u^2 - v^2)h(u)g(v) & = & 0 \nonumber
\end{eqnarray}
and
\begin{eqnarray}
((u^2 - v^2)(u - v) - (u + v)(u - v - 2))(h(v)f(u) - f(v)h(u)) +  2(u^2 - v^2)(u - v)h(v)g(u) &&  \\
+ 2(u^2 - v^2)h(v)f(u) - 2(u^2 - v^2)(u + v)f(v)g(u) + 4(u - v)(u + v + 1)(u + v - 2)g(u)g(v) && \nonumber \\
+ (u - v)((u + v)^2 - (u + v - 2))h(u)h(v) + (u - v)((u + v - 2) - (u + v)(u + v + 2))f(v)f(u) & = & 0. \nonumber
\end{eqnarray}

Finally, the case with all external $SO(6)$ indices in the $SO(4)$ gives us the two equations
\begin{eqnarray}
(u + v)(f(v)g(u) - f(u)g(v)) - (u - v)(f(v)g(u) + f(u)g(v)) && \nonumber \\
- 2(u^2 - v^2)g(u)g(v) & = & 0
\end{eqnarray}
and
\begin{eqnarray}
((u^2 - v^2)(u - v - 2) - (u - v)(u + v - 2))(f(v)g(u) + f(u)g(v)) + 4(u^2 - v^2)f(u)g(v) &&  \\
((u + v)(u - v - 2) - (u^2 - v^2)(u + v - 2))(f(v)g(u) - f(u)g(v)) + 2(u^2 - v^2)h(u)g(v) && \nonumber \\
+ 2(u^2 - v^2)(u^2 - v^2 - 2u + 2)g(u)g(v) & = & 0. \nonumber
\end{eqnarray}

These equations determine the functions $h(u), f(u), g(u)$ (up to normalization) to be
\begin{eqnarray}
h(u) & = & 1 - u^2 \nonumber \\
g(u) & = & -u \nonumber \\
f(u) & = & 1 + u^2.
\end{eqnarray}

This means that we have
\begin{eqnarray}
(\mathcal{K}^{-})^{b J}_{a I}(u) & = & (1 + u^2)\delta^{b}_{a}\delta^{J}_{I} - u\bar{W}^{J}_{ac}W_{I}^{cb} \nonumber \\
(\mathcal{K}^{-})^{b Y}_{a X}(u) & = & (1 - u^2)\delta^{b}_{a}\delta^{Y}_{X} \nonumber \\
(\mathcal{K}^{-})^{b Y}_{a I}(u) & = & 0 \nonumber \\
(\mathcal{K}^{-})^{b J}_{a X}(u) & = & 0.
\end{eqnarray}

In addition, it can be shown that if $\mathcal{K}^{-}_{a}(u)$ satisfies its BYB, then $\mathcal{K}_{a}^{+}(u) = \mathcal{K}_{a}^{- t_{a}}(2 - u)$ satisfies the other BYB, so that we have
\begin{eqnarray}
(\mathcal{K}^{+})^{b J}_{a I}(u) & = & (5 - 4u + u^2)\delta^{b}_{a}\delta^{J}_{I} + (u - 2)\bar{W}_{I ac}W^{J cb} \nonumber \\
(\mathcal{K}^{+})^{b Y}_{a X}(u) & = & (-3 + 4u - u^2)\delta^{b}_{a}\delta^{Y}_{X} \nonumber \\
(\mathcal{K}^{+})^{b Y}_{a I}(u) & = & 0 \nonumber \\
(\mathcal{K}^{+})^{b J}_{a X}(u) & = & 0.
\end{eqnarray}

Now, if we formed a transfer matrix $\hat{t}(u)$ from these two operators, each of the ends would act on the same representation of $SU(2)$.  However, our operators have one end in the fundamental, and one in the anti-fundamental.  Therefore, we create
\begin{equation}
(\widetilde{\mathcal{K}}^{-})^{a j}_{b i} = \epsilon^{ac}(\mathcal{K}^{-})^{d j}_{c i}\epsilon_{d b}
\end{equation}
(which must also satisfy the BYB) and we use this object in the creation of our transfer matrix:
\begin{eqnarray}
(\widetilde{\mathcal{K}}^{-})^{a J}_{b I}(u) & = & (1 + u^2)\delta^{a}_{b}\delta^{J}_{I} - u\bar{W}^{J}_{bc}W_{I}^{ca} \nonumber \\
(\widetilde{\mathcal{K}}^{-})^{a Y}_{b X}(u) & = & (1 - u^2)\delta^{a}_{b}\delta^{Y}_{X} \nonumber \\
(\widetilde{\mathcal{K}}^{-})^{b Y}_{a I}(u) & = & 0 \nonumber \\
(\widetilde{\mathcal{K}}^{-})^{b J}_{a X}(u) & = & 0.
\end{eqnarray}
Using these objects, we can now expand our transfer matrix, and we find that 
\begin{equation}
\hat{t}_{0} = \hat{t}(0) = 6I
\end{equation}
and
\begin{equation}
\hat{t}_{1} = \frac{d\hat{t}}{du}(0) = -7I - 6R_{1} - 6R_{L} - 6\sum_{n = 1}^{L - 2} (2P_{n n+1} + I_{n n+1} - K_{n n+1})
\end{equation}
where $R_{1}$ and $R_{L}$ are as defined in (\ref{eq:Rdef}).  Thus, we find that the anomalous dimension matrix from the previous section can, indeed, be written as a linear combination of these operators
\begin{equation}
\Lambda = \frac{g^2 N}{16 \pi^2}\left[\frac{-1}{6}\hat{t}_{1} - \frac{1}{6}\left(\frac{13}{6} + 3L\right)\hat{t}_{0}\right]
\end{equation}
and thus that the eigenvalues of this matrix can be found using a Bethe ansatz.

\section{The Bethe Ansatz}

Having demonstrated that our spin chain is integrable, we now want to find the eigenstates and eigenvalues of the anomalous dimension matrix, using the Bethe ansatz.  The basic idea is to start with a reference state of maximal charge under one $SO(2)$ subgroup of $SO(4) \times SO(2)$, and then change the charges of some of the fields in the operators, creating ``impurities.''  These impurities form spin waves that have momenta quantized by the Bethe ansatz
\begin{equation}
e^{ip_{i}L} = \prod_{j \ne i} S_{ji}(p_{j}, p_{i})
\end{equation}
where $p_{i}$ label the momenta of the spin waves, and $S_{ij}(p_{i}, p_{j})$ is the S-matrix for the scattering of two spin waves.  The general formula for the Bethe ansatz for a spin chain with sites in a given representation of a given Lie algebra is \cite{MZ}
\begin{eqnarray}
\label{LieBA}
\left(\frac{u_{q,i}+i\vec\alpha_q\cdot \vec w/2}{u_{q,i}-i\vec\alpha_q\cdot \vec w/2}\right)^L = \prod_{j\ne i}^{n_q}\frac{u_{q,i}-u_{q,j}+i\vec\alpha_q\cdot\vec\alpha_q/2}{u_{q,i}-u_{q,j}-i\vec\alpha_q\cdot\vec\alpha_q/2}
\prod_{q'\ne q}\prod_{j}^{n_{q'}}\frac{u_{q,i}-u_{q',j}+i\vec\alpha_q\cdot\vec\alpha_{q'}/2}{u_{q,i}-u_{q',j}-i\vec\alpha_q\cdot\vec\alpha_{q'}/2} \,.
\end{eqnarray}
Here the $u_{q,i}$ are parameters characterizing excitations,
taking the place of the $p_i$; $i$ labels the excitation as
before, while $q$ reflects the fact that the excitation can be
associated to any of the simple roots $\vec\alpha_q$ of the
algebra. $\vec{w}$ is the highest weight vector of the
representation of the group that lives at each site.  For a
fundamental at each site, we will have $\vec{w} = \vec{w}^1$, the
first fundamental weight, which has the inner product with simple
roots $\vec\alpha_q \cdot \vec{w}^1 = \delta_q^1$.

Note that the relationship between the spin wave momentum $k$ and the new parameter $u$ is \cite{MZ}
\begin{eqnarray}
p(u_{1,i})= -i \log { u_{1,i} + i/2 \over u_{1,i} - i/2} \,.
\end{eqnarray}

In this paper we restrict ourselves to operators with a reference state of maximal charge under one $SO(2)$, with spin waves all of the same definite charge under another $SO(2)$.  In this case the Bethe ansatz for the closed spin chain simplifies to \cite{Beis}
\begin{equation}
\left(\frac{u_{j} + i/2}{u_{j} - i/2}\right)^{L} = \prod_{k \ne j}^{J} \frac{u_{j} - u_{k} + i}{u_{j} - u_{k} - i},
\end{equation}
and the spin-wave S-matrix is
\begin{equation}
S_{kj}(p_{k}, p_{j}) = \frac{u_{j} - u_{k} + i}{u_{j} - u_{k} - i}.
\end{equation}

We know that the generalization of this Bethe ansatz for open spin chains should be
\begin{equation}\label{eq:BBA}
e^{2ip_{i}L} = \mathcal{B}_{1}(p_{i})\mathcal{B}_{2}(p_{i})\prod_{j \ne i} (S_{ji}(p_{j}, p_{i})S_{ji}(-p_{j}, p_{i}))
\end{equation}
where the $\mathcal{B}_{1,2}$ are possible phases that a spin wave could obtain from reflecting off the boundary.

As was pointed out in \cite{ME}, since we already know the S-matrix from the closed spin chains, all we need to determine are the factors $\mathcal{B}_{1,2}$.  Furthermore, these can be found by considering states of only a single impurity, since only one impurity interacts with the boundary at a time.

\subsection{The Bethe Reference State}

We begin with our reference state, which is
\begin{equation}
\bar{Q}_{2}X^{L}Q^{1}.
\end{equation}
If we say that $J_{1}$, $J_{2}$, and $J_{3}$  are charges under rotations of the three $SO(2)$ subgroups corresponding to $X$, $Y$, and $Z$, then this state has charges
\begin{equation}
(J_{1}, J_{2}, J_{3}) = (L+1, 1, 0).
\end{equation}
This is somewhat unusual for a Bethe reference state, which is usually constructed to have charge under only one $SO(2)$; however, it is (as we discussed earlier) the operator of this type with maximal charge $J_{1}$, and it does have vanishing anomalous dimension.

\subsection{Single Impurities of Type $Z$ or $\bar{Z}$}

We now to consider operators where one of the $X$ fields in the reference state is replaced with a $Z$ field spin wave, creating operators of charge
\begin{equation}
(J_{1}, J_{2}, J_{3}) = (L, 1, 1).
\end{equation}
(Operators with a $\bar{Z}$ impurity act identically).  The operators in this sector are of the form
\begin{equation}
|Z(x)\rangle = \bar{Q}_{2}X^{x-1}ZX^{L - x}Q^{1},
\end{equation}
and we take a standard Bethe ansatz linear combination of these
\begin{equation}
|Z(p)\rangle = \sum_{x = 1}^{L} (A(p)e^{ipx} + \tilde{A}e^{-ipx})|Z(x)\rangle
\end{equation}
to be our proposed eigenstate.  Using our operator (\ref{eq:gamma}), we find that
\begin{eqnarray}
\Gamma|Z(x)\rangle & = & -\lambda\left[2|Z(x - 1)\rangle + 2|Z(x + 1)\rangle - 4|Z(x)\rangle\right]  \ \ \ \ (x \ne 1, L) \nonumber \\
\Gamma|Z(1)\rangle & = & \lambda\left[4|Z(1)\rangle - 2|Z(2)\rangle\right] \nonumber \\
\Gamma|Z(L - 1)\rangle & = & \lambda\left[4|Z(L)\rangle - 2|Z(L-1)\rangle\right].
\end{eqnarray}
We then use this to calculate
\begin{eqnarray}
\Gamma|Z(p)\rangle & = & 4\lambda(1 - \cos p)\sum_{x = 2}^{L - 1} (A(p)e^{ipx} + \tilde{A}(p)e^{-ipx})|Z(x)\rangle \nonumber \\
&& + \lambda\left(A(p)e^{ip}(4 - 2e^{ip}) + \tilde{A}(p)e^{-ip}(4 - 2e^{-ip})\right)|Z(1)\rangle \nonumber \\
&& + \lambda\left(A(p)e^{ipL}(4 - 2e^{-ip}) + \tilde{A}(p)e^{-ipL}(4 - 2e^{ip})\right)|Z(L)\rangle.
\end{eqnarray}
Notice that for the terms of $|Z(p)\rangle$ with $x = 2, ..., L-1$, we already have what we want, so that $4\lambda(1 - \cos p)$ must be the eigenvalue of $|Z(p)\rangle$ under operator $\Gamma$.  In order for the operator to be an eigenstate, we then demand that
\begin{equation}
4(1 - \cos p)(A(p)e^{ip} + \tilde{A}(p)e^{-ip}) = (4 - 2e^{ip})A(p)e^{ip} + (4 - 2e^{-ip})\tilde{A}(p)e^{-ip},
\end{equation}
which gives
\begin{equation}
A(p) = -\tilde{A}(p)
\end{equation}
and
\begin{equation}
4(1 - \cos p)(A(p)e^{ipL} + \tilde{A}(p)e^{-ipL}) = (4 - 2e^{-ip})A(p)e^{ipL} + (4 - 2e^{ip})\tilde{A}(p)e^{-ipL},
\end{equation}
which (together with the above result) gives
\begin{equation}
e^{2ip(L + 1)} = 1.
\end{equation}
This is then our Bethe equation for a single impurity of this type.  Notice that the $Z$ impurity ``sees'' a spin chain of effective length $L + 1$, and that the boundary terms are trivial:
\begin{equation}
\mathcal{B}_{1,2} = 1.
\end{equation}
In total, we write that the spin wave momentum is quantized as
\begin{equation}
p = \frac{\pi n}{L + 1}, \ \ \ n \in \mathbb{Z},
\end{equation}
the eigenstate is written as
\begin{equation}
|Z(p)\rangle = \sum_{x = 1}^{L} \sin \frac{\pi n x}{L + 1}|Z(x)
\end{equation}
(so that the spin wave does, indeed, obey ``Dirichlet boundary conditions,'' as one would expect of an impurity that corresponds to a direction perpendicular to the 7-brane), and the anomalous dimension of this operator is given by
\begin{equation}
\Gamma|Z(p)\rangle = 4\lambda\left(1 - \cos \frac{\pi n}{L + 1}\right)|Z(p)\rangle.
\end{equation}

\subsection{Impurities of Type $Y$}

Next we consider states with a single impurity of type $Y$, so that the states have charge
\begin{equation}
(J_{1}, J_{2}, J_{3}) = (L, 2, 0).
\end{equation}
Notice that even though these impurities are charged under the $SO(4)$ under which the boundary terms are charged, this sector does not contain any operators where the boundary fields are ``flipped,'' for example, operators like $\bar{Q}_{1}XXX\cdots$.  Therefore, the calculations here are very similar to the previous ones.  We label the operators in this sector
\begin{equation}
|Y(x)\rangle = \bar{Q}_{2}X^{x - 1}YX^{L - x}Q^{1}
\end{equation}
and look for an eigenstate of the form
\begin{equation}
|Y(p)\rangle = \sum_{x = 1}^{L} (B(p)e^{ipx} + \tilde{B}(p)e^{-ipx})|Y(x)\rangle.
\end{equation}
We quickly find that
\begin{eqnarray}
\Gamma|Y(x)\rangle & = & \lambda\left[4|Y(x)\rangle - 2|Y(x - 1)\rangle - 2|Y(x+1)\rangle\right], \ \ \ x \ne 1, L \nonumber \\
\Gamma|Y(1)\rangle & = & \lambda\left[2|Y(1)\rangle - 2|Y(2)\rangle\right] \nonumber \\
\Gamma|Y(L)\rangle & = & \lambda\left[2|Y(L)\rangle - 2|Y(L-1)\rangle\right],
\end{eqnarray}
and this leads to the equation
\begin{eqnarray}
\Gamma|Y(p)\rangle & = & 4\lambda(1 - \cos p)\sum_{x = 2}^{L - 1}  (B(p)e^{ipx} + \tilde{B}e^{-ipx})|Y(x)\rangle \nonumber \\
&& + \lambda\left((2 - 2e^{ip})B(p)e^{ip} + (2 - 2e^{-ip})\tilde{B}(p)e^{-ip}\right)|Y(1)\rangle \nonumber \\
&& + \lambda\left((2 - 2e^{-ip})B(p)e^{ipL} + (2 - 2^{ip})\tilde{B}(p)e^{-ipL}\right)|Y(L)\rangle.
\end{eqnarray}
From here, we can see that the eigenvalue of this state must be $4\lambda(1 - \cos p)$, and in order for the state to be an eigenstate we need to have
\begin{equation}
4(1 - \cos p)(B(p)e^{ip} + \tilde{B}(p)e^{-ip}) = (2 -2e^{ip})B(p)e^{ip} + (2 - 2e^{-ip})\tilde{B}(p)e^{-ip},
\end{equation}
which gives us
\begin{equation}
\tilde{B}(p) = e^{ip}B(p)
\end{equation}
and
\begin{equation}
4(1 - \cos p)(B(p)e^{ipL} + \tilde{B}(p)e^{-ipL}) = (2 - 2e^{-ip})B(p)e^{ipL} + (2 - 2e^{-ip})\tilde{B}(p)e^{-ipL},
\end{equation}
which (together with the previous result) gives us
\begin{equation}
e^{2ipL} = 1.
\end{equation}
Thus, again we have that
\begin{equation}
\mathcal{B}_{1,2} = 1.
\end{equation}
This type of impurity apparently ``sees'' a spin chain of length $L$, and the momentum of the spin wave is quantized as
\begin{equation}
p = \frac{n\pi}{L}.
\end{equation}
The eigenstates are 
\begin{equation}
|Y(p)\rangle = \sum_{x = 1}^{L} \cos \frac{n\pi(x - 1/2)}{L} |Y(x)\rangle
\end{equation}
(and here the spin wave obeys ``Neumann boundary conditions,'' just as it should), and their eigenvalues are
\begin{equation}
\Gamma|Y(p)\rangle = 4\lambda\left(1 - \cos \frac{n\pi}{L}\right)|Y(k)\rangle.
\end{equation}

\subsection{Impurities of Type $\bar{Y}$}

Finally, we consider eigenstates in the sector with charges
\begin{equation}
(J_{1}, J_{2}, J_{3}) = (L, 0, 0).
\end{equation}
This sector contains operators of the type
\begin{equation}
|\bar{Y}(x)\rangle = \bar{Q}_{2}X^{x -1}\bar{Y}X^{L - x}Q^{1},
\end{equation}
but it also contains the operators
\begin{equation}
|\bar{Q}_{1}\rangle = \bar{Q}_{1}X^{L}Q^{1}, \ \ \ \ \ |Q^{2}\rangle = \bar{Q}_{2}X^{L}Q^{2}.
\end{equation}
Thus, our proposed eigenstates will have the form
\begin{equation}
|\bar{Y}(p)\rangle = D(p)|\bar{Q}_{1}\rangle + E(p)|Q^{2}\rangle + \sum_{x = 1}^{L} (C(p)e^{ipx} + \tilde{C}(p)e^{-ipx})|\bar{Y}(x)\rangle.
\end{equation}
We find that
\begin{eqnarray}
\Gamma|\bar{Y}(x)\rangle & = & \lambda\left[4|\bar{Y}(x)\rangle - 2|\bar{Y}(x-1)\rangle - 2|\bar{Y}(x+1)\rangle\right], \ \ \ \ x \ne 1, L \nonumber \\
\Gamma|\bar{Y}(1)\rangle & = & \lambda\left[4|\bar{Y}(1)\rangle + 2|\bar{Q}_{1}\rangle - 2|\bar{Y}(2)\rangle\right] \nonumber \\
\Gamma|\bar{Y}(L)\rangle & = & \lambda\left[4|\bar{Y}(L)\rangle - 2|Q^{2}\rangle - 2|\bar{Y}(L-1)\rangle\right] \nonumber \\
\Gamma|\bar{Q}_{1}\rangle & = & \lambda\left[2|\bar{Q}_{1}\rangle + 2|\bar{Y}(1)\rangle\right] \nonumber \\
\Gamma|Q^{2}\rangle & = & \lambda\left[2|Q^{2}\rangle - 2|\bar{Y}(L)\rangle\right],
\end{eqnarray}
which gives us
\begin{eqnarray}
\Gamma|\bar{Y}(p)\rangle & = & 4\lambda(1 - \cos p)\sum_{x = 2}^{L-1} (C(p)e^{ipx} + \tilde{C}(p)e^{-ipx})|\bar{Y}(x)\rangle \nonumber \\
&& + \lambda\left[(4 - 2e^{ip})C(p)e^{ip} + (4 - e^{-ip})\tilde{C}(p)e^{-ip} + 2D(p)\right]|\bar{Y}(1)\rangle \nonumber \\
&& + \lambda\left[(4 - 2e^{-ip})C(p)e^{ipL} + (4 - e^{ip})\tilde{C}(p)e^{-ipL} - 2E(p)\right]|\bar{Y}(L)\rangle \nonumber \\
&& + \lambda\left[2C(p)e^{ip} + 2\tilde{C}(p)e^{-ip} + 2D(p)\right]|\bar{Q}_{1}\rangle \nonumber \\
&& + \lambda\left[-2C(p)e^{ipL} - 2\tilde{C}(p)e^{-ipL} + 2E(p)\right]|Q^{2}\rangle.
\end{eqnarray}
From the first term, we see that the eigenvalue will again be $4\lambda(1 - \cos p)$, and to ensure that the state is an eigenstate we then need that
\begin{equation}
4(1 - \cos p)(C(p)e^{ip} + \tilde{C}(p)e^{-ip}) = (4 - 2e^{ip})e^{ip}C(p) + (4 - 2e^{-ip})e^{-ip}\tilde{C}(p) + 2D(p),
\end{equation}
which gives
\begin{equation}
D(p) = -C(p) - \tilde{C}(p)
\end{equation}
and
\begin{equation}
4(1 - \cos p)(C(p)e^{ipL} + \tilde{C}(p)e^{-ipL}) = (4 - 2e^{-ip})C(p)e^{ipL} + (4 - 2e^{ip})\tilde{C}(p)e^{-ipL} - 2E(p),
\end{equation}
which gives
\begin{equation}
E(p) = C(p)e^{ip(L+1)} + \tilde{C}(p)e^{-ip(L+1)}.
\end{equation}
(Note that the boundary terms are now acting like extra sites in the spin chain.)  We also need
\begin{equation}
4(1 - \cos p)D(p) = 2C(p)e^{ip} + 2\tilde{C}(p)e^{-ip} + 2D(p),
\end{equation}
which, together with the previous result, yields
\begin{equation}
\tilde{C}(p) = e^{ip}C(p)
\end{equation}
and
\begin{equation}
4(1 - \cos p)E(p) = 2E(p) - 2C(p)e^{ipL} - 2\tilde{C}(p)e^{-ipL}.
\end{equation}
This, together with the previous results, yields
\begin{equation}
e^{2ip(L+2)} = 1,
\end{equation}
which is our spin-wave momentum quantization condition for this type of impurity.  Again, we have seen that the boundary interaction is trivial:
\begin{equation}
\mathcal{B}_{1,2} = 1.
\end{equation}
Our spin wave momentum is now quantized according to
\begin{equation}
p = \frac{n\pi}{L+2}
\end{equation}
because this type of impurity ``sees'' an effective length of $L+2$.  With the definitions
\begin{equation}
|\bar{Y}(0)\rangle \equiv -|\bar{Q}_{1}\rangle, \ \ \ \ \ |\bar{Y}(L+1)\rangle \equiv |Q^{2}\rangle,
\end{equation}
our eigenstates are
\begin{equation}
|\bar{Y}(p)\rangle = \sum_{x = 0}^{L+1} \cos \frac{n\pi(x + 1/2)}{L+2}|\bar{Y}(x)\rangle
\end{equation}
(again with Neumann boundary conditions for the spin waves), and their eigenvalues are
\begin{equation}
\Gamma|\bar{Y}(p)\rangle = 4\lambda\left(1 - \cos \frac{n\pi}{L+2}\right)|\bar{Y}(p)\rangle.
\end{equation}

There are three results in this section that we would like to highlight.  First, notice that for all three types of impurities, the spin waves reflect trivially off the boundary.  This is consistent with the results of \cite{ME}.  It is probable that any open spin chains in a CFT whose boundary terms are fundamental matter created by the insertion of a simple D-brane into the AdS$_{5}\times$S$^{5}$ background will behave like this.  These types of open spin chains are related to closed spin chains by simple ``doubling tricks,'' just as the open strings are related to the closed strings.  It would be interesting to know whether any kind of fundamental matter can lead to nontrivial reflection terms and, if so, what kind of string theory solution they correspond to.

Second, the spin waves that should correspond to motion of the open string parallel to the D-brane satisfy Neumann boundary conditions, while those that correspond to motion perpendicular to the D-brane satisfy Dirichlet boundary conditions.  Again, this reinforces the previous results of \cite{ME} as further evidence of the AdS/CFT correspondence.  

Finally, consider the effective lengths of the spin chains, as seen from the quantization conditions of the spin-wave momenta.  In \cite{ME}, where the R-symmetry was broken to $SO(3)_{V} \times SO(3)_{H}$ with the boundary terms charged under $SO(3)_{H}$, DeWolfe and Mann showed that impurities charged under the $SO(3)_{V}$ ``saw'' an effective length of one unit less than the effective length experienced by $SO(3)_{H}$ impurities.  The authors argued that this was because the impurities in $SO(3)_{H}$ could interact with the boundaries, which then, together, acted like an extra site in the spin chain.  Here, the story is clearly more complicated.  We have $Z$ or $\bar{Z}$ impurities experiencing effective length $L+1$, $Y$ impurities experiencing effective length $L$, and $\bar{Y}$ impurities experiencing effective length $L+2$.  This effective length matters when we consider the use of equation (\ref{eq:BBA}) in connection with the ``doubling trick,'' in the next section.

\section{Application of the ``Doubling Trick'' and Relation to Spinning Strings}

In \cite{OSC/OSS}, Chen, Wang, and Wu showed that open spin chains could be directly related to closed spin chains, and, similarly, open spinning strings could be related to closed spinning strings.  These two relations then were shown to imply that if the closed spin chains and closed spinning strings satisfied the AdS/CFT correspondence, so did the open spin chains and open spinning strings.  Here, we apply these arguments to the particular open spin chains being studied in this paper.

Consider first an open spin chain with a large number of impurities of type $Z$ (or $\bar{Z}$).  This spin chain will have charges
\begin{equation}
(J_1, J_2, J_3) = (j_1+1, 1, j_3)
\end{equation}
where $j_1$ and $j_3$ are the number of $X$ and $Z$ fields in the interior of the operator.  This spin chain has roots that satisfy the equation
\begin{equation}\label{eq:rootsZ}
\left(\frac{u_j + i/2}{u_j - i/2}\right)^{2(1+j_1+j_3)} = \prod_{k \ne j}^{j_3}\frac{u_j - u_k + i}{u_j - u_k - i}\frac{u_j + u_k + i}{u_j + u_k - i}
\end{equation}
and has anomalous dimension
\begin{equation}
\gamma_{o}(j_1+1, 1, j_3) = \sum_{j}^{j_3} 4\lambda(1 - \cos p(u_j)) \approx \sum_{j}^{j_3} 2\lambda p(u_j)^2.
\end{equation}
However, equation (\ref{eq:rootsZ}) can be directly related to the equation for a closed spin chain of length $2(1+j_1+j_3) + 1$ with $2j_3$ symmetrically distributed roots \footnote{This equation includes a small correction to equation (24) in reference \cite{OSC/OSS}.}
\begin{equation}
\left(\frac{u_j + i/2}{u_j - i/2}\right)^{2(1+j_1+j_3) + 1} = \prod_{k \ne j}^{2j_3}\frac{u_j - u_k + i}{u_j - u_k - i}. 
\end{equation}
Thus we can relate the two anomalous dimensions as
\begin{equation}
\gamma_{o}(j_1+1, 1, j_3) = \frac{1}{2}\gamma_{c}^{\mbox{sym}}(2j_1+3, 0, 2j_3).
\end{equation}
 
Notice that the related closed spin chain carries no evidence of the $J_2$ charge carried by the open chain; this is an effect that disappears in the thermodynamic limit, but should still somehow be explained on the string side eventually, through D-brane effects.  We can write this relationship in terms of the actual dimension of the operators, which can then be related to the energies of open and closed strings, at least to lowest order in $\lambda/J^2$.  Doing so gives us
\begin{equation}
\Delta_{o}(j_1+1,1,j_3) = \frac{1}{2}\Delta_{c}(2j_1+3, 0, 2j_3) + \frac{1}{2}
\end{equation}
and then
\begin{equation}
E_{o}(j_1+1, 1, j_3) = \frac{1}{2}E_{c}(2j_1+3, 0, 2j_3) + \frac{1}{2}.
\end{equation}
This gives results that will differ from those in \cite{OSC/OSS} by terms of order $1/J_i$, which vanish in the semi-classical limit.  

By similar logic, we can work out the relation between open spin chains with $Y$ impurities and a closed spin chain to be
\begin{equation}
\gamma_{o}(j_1+1, j_2+1, 0) = \frac{1}{2}\gamma_{c}(2j_1+1, 2j_2, 0)
\end{equation}
which gives the relationship between energies of open and closed strings
\begin{equation}
E_{o}(j_1+1, j_2+1, 0) = \frac{1}{2}E_{c}(2j_1+1, 2j_2, 0) + \frac{3}{2}.
\end{equation}

We can do the same for the open spin chain with $\bar{Y}$ impurities and we find the relationship
\begin{equation}
\gamma_{o}(j_1+1, 1 - j_2, 0) = \frac{1}{2}\gamma_{c}^{\mbox{sym}}(5 + 2j_1 , -2j_2, 0).
\end{equation}
which gives
\begin{equation}
E_{o}(j_1+1, 1 - j_2, 0) = \frac{1}{2}E_{c}(5+2j_1, -2j_2, 0) - \frac{1}{2}.
\end{equation}
This is a more interesting result than those above, because it differs from the expected doubling relation by terms of order $1$, rather than of order $1/J_i$.  This is due to the fact that the impurities carry opposite R-charge from the fundamental matter on the ends of the spin chain.   The closed spin chain that we relate to the open spin chain does not really incorporate the $Y$ R-charge from the fundamental matter at the ends of the spin chain, so it is a simpler object.

Finally, for comparison, we can carry this procedure out on the spin chains studied in \cite{ME}, where the fundamental matter is in a dCFT.  Using impurities that do not interact with the boundaries, (analogous to our $Z$ impurities), we would find the anomalous dimension relation
\begin{equation}
\gamma_{o}^{\mbox{dCFT}}(1+j_1, 0, j_3) = \frac{1}{2}\gamma_{c}^{\mbox{sym}}(2j_1+1, 0, 2j_3).
\end{equation}
and the energy relation
\begin{equation}
E_{o}^{\mbox{dCFT}}(1+j_1, 0, j_3) = \frac{1}{2} + \frac{1}{2}E_{c}(2j_1+1, 0, 2j_3)
\end{equation}
which again differs from the expected result by a term of order $1/J_{i}$.

Notice that in all these cases, the anomalous dimension of the open spin chain is directly related to the anomalous dimension of a closed spin chain.  However, the associated closed spin chain is simpler than the open spin chain.  The closed spin chain is determined completely by the impurity spin waves and the effective length that these spin waves ``see''.  However, this does not capture the full charge structure of the open spin chain, and so some information is lost in the relationship.  Specifically, the closed spin chain is relatively unaffected by the details of the ends of the open spin chain, even when the length of the spin chain is small.

\section{Conclusions and Open Questions}

In this paper we have found one more supersymmetric gauge theory with fundamental scalar matter that has one-loop integrability.  We solved the Bethe ansatz and determined the spectrum, confirming that reflections of spin waves off the ends of the spin chain correspond directly to boundary conditions of an open string stuck to a D7-brane.  We also used a ``doubling trick'' to relate open spin chain excitations with anomalous dimension $\gamma_o$ to closed spin chain excitations with anomalous dimension $2\gamma_o$.

However, we found that this doubling trick exhibited unexpected results.  The charges of the closed spin chain were not just ``double'' that of the associated open spin chain.  This is because the closed spin chain is not sensitive to all of the details of the boundaries of the open spin chain, even when the spin chain is short.  The boundary terms interact differently with different types of impurities, but some of this information is lost when relating the open spin chain to the closed spin chain.  After applying the AdS/CFT correspondence to this, we found an implied relationship between the energies of open and closed strings, but because of the above issues, this relationship differed from the expected doubling trick in all cases by terms of order $\mathcal{O}(1/J)$, and in one case by a term of order $\mathcal{O}(1)$.  The terms of order $\mathcal{O}(1/J)$ represent finite size effects that clearly vanish in the semiclassical approximations that have been studied.  The term of order $\mathcal{O}(1)$ arises because the closed string state associated with the open string state is not the obvious one.  This suggests that care needs to be exercised in using the doubling trick for open strings and closed strings in the AdS$_5\times$S$^5$, so that the appropriate states are identified.

This work suggests that the boundary conditions of open strings can give corrections to the $\lambda/J^2$ term in the energies of these open strings of order $1/J$, (that is, there is a correction to the energy of the string of order $\lambda/J^3$).  Very little work has been done to separate the quantum corrections to open strings from pure finite size corrections.  It would be interesting to develop techniques to study these types of corrections, in order to verify the relations we found here using the Bethe ansatz techniques.  It would also be useful to construct the open string related to our $\bar{Y}$-impurity spin chain, and see if we can reproduce the $\mathcal{O}(1)$ discrepancy in the doubling trick, and better understand it from the string side.

\section{Acknowledgments}

We would like to thank Oliver DeWolfe, Joe Polchinski, and David Berenstein for many useful and illuminating discussions.   This material is based upon work supported by the National Science  Foundation under grant  PHY00-98395.  Any opinions, findings, and conclusions or recommendations expressed in  this material are those of the authors and do not necessarily reflect the views of the National Science Foundation. The work of N.M. \ was also supported by a National Defense Science and Engineering Graduate Fellowship.

\appendix

\section{Appendix: Field Theory Conventions and Details}

The matter of the theory and its representation under the R-symmetry are summarized in table \ref{N=2table} \cite{mesons}, with the $\mathcal{N} = 4$ information presented for comparison in table \ref{N=4table}.  

\begin{table}
\begin{tabular}{|c|c|c|c|c|}
\hline
field & $SO(4) = SU(2)_{L} \times SU(2)_{R}$ & $SO(2) = U(1)$ & $SU(N)$ &  $SO(3,1)$ \\
\hline
\hline
$\phi_{I}$ & fundamental of & neutral & adjoint & scalars \\
& $SO(4)$  $I= 1, ... 4$ & & & \\
\hline
$Z = \phi_{5} + i\phi_{6}$ & neutral & $+1$ charge & adjoint & complex \\
& & under $U(1)$ & & scalars \\
\hline
$\Lambda_{a}^{\alpha}$ & $a = 1,2$,   $(0,1/2)$ & $+1/2$ charge & adjoint & Weyl spinors, \\
& of $SU(2)_{L} \times SU(2)_{R}$ & under $U(1)$ & & $\alpha = 1,2$ \\
\hline
$\Theta_{\bar{a}}^{\alpha}$ & $\bar{a} = 1,2$,   $(1/2, 0)$ & $-1/2$ charge & adjoint & Weyl spinors, \\
& of $SU(2)_{L} \times SU(2)_{R}$ & under $U(1)$ & & $\alpha = 1,2$ \\
\hline
$A^{\mu}$ & neutral & neutral & adjoint & vector, \\
&&&& $\mu = 0, ..., 3$ \\
\hline
$Q^{a}$ & $a = 1,2$,   $(0, -1/2)$ & neutral & fundamental & complex \\
 & of $SU(2)_{L} \times SU(2)_{R}$ & & & scalars \\
 \hline
 $\chi^{\alpha}$ & neutral & $-1/2$ charge & fundamental & Weyl spinors, \\
 & & under $U(1)$ & & $\alpha = 1,2$ \\
 \hline
 $\pi^{\alpha}$ & neutral & $-1/2$ charge & anti- & Weyl spinors, \\
 & & under $U(1)$ & fundamental & $\alpha = 1,2$ \\
 \hline
 \end{tabular}
\caption{\label{N=2table} A table of the fields in our $\mathcal{N} = 2$ theory, with their representations under the global symmetry groups, the gauge symmetry group, and the Lorentz symmetry group.  Gauge indices are suppressed.}
\end{table}

\begin{table}
\begin{tabular}{|c|c|c|c|}
\hline
field & $SO(6) = SU(4)$ & $SU(N)$ & $SO(3,1)$ \\
\hline
\hline
$\phi_{i}$ & fundamental of & adjoint & scalars \\
& $SO(6)$   $ i = 1, ..., 6$ & & \\
\hline
$\psi_{A}^{\alpha}$ & fundamental of & adjoint & Weyl spinors, \\
& $SU(4)$   $A = 1, ..., 4$ && $\alpha = 1,2$ \\
\hline
$A^{\mu}$ & neutral & adjoint & vector, \\
&&& $\mu = 0, ..., 3$ \\
\hline
\end{tabular}
\caption{\label{N=4table} For reference, a table of the fields in the $N = 4$ theory, with their representations under the various symmetry groups.}
\end{table}

The $C_{i}^{AB}$ are Clebsch-Gordan matrices that translate between the $4$ and the $6$ representation of $SU(4)$.  They are as follows:

\begin{equation}
C_{1} = \left(\begin{array}{cccc} 0 & 0 & i & 0 \\ 0 & 0 & 0 & -i \\ -i & 0 & 0 & 0 \\ 0 & i & 0 & 0 \end{array}\right), \ \ \ \ C_{2} = \left(\begin{array}{cccc} 0 & 0 & 0 & -1 \\ 0 & 0 & 1 & 0 \\ 0 & -1 & 0 & 0 \\ 1 & 0 & 0 & 0 \end{array}\right)
\end{equation}
\begin{displaymath}
C_{3} = \left(\begin{array}{cccc} 0 & 0 & i & 0 \\ 0 & 0 & 0 & -i \\ -i & 0 & 0 & 0 \\ 0 & i & 0 & 0 \end{array}\right), \ \ \ \ C_{4} = \left(\begin{array}{cccc} 0 & 0 & -1 & 0 \\ 0 & 0 & 0 & -1 \\ 1 & 0 & 0 & 0 \\ 0 & 1 & 0 & 0 \end{array}\right)
\end{displaymath}
\begin{displaymath}
C_{5} = \left(\begin{array}{cccc} 0 & -i & 0 & 0 \\ i & 0 & 0 & 0 \\ 0 & 0 & 0 & -i \\ 0 & 0 & i & 0 \end{array}\right), \ \ \ \ C_{6} = \left(\begin{array}{cccc} 0 & 1 & 0 & 0 \\ -1 & 0 & 0 & 0 \\ 0 & 0 & 0 & -1 \\ 0 & 0 & 1 & 0 \end{array}\right).
\end{displaymath}

And the $\bar{C}^{i}_{AB}$ are the Hermitian conjugates of these.  The $W_{I}^{\bar{a}b}$ translate between the $4$ and the $(1/2, 1/2)$ of $SO(4)$, and are as follows:

\begin{equation}
W_{1} = \left(\begin{array}{cc} 0 & -1 \\ -1 & 0 \end{array}\right), \ \ \ \ W_{2} = \left(\begin{array}{cc} 0 & i \\ -i & 0 \end{array}\right)
\end{equation}
\begin{displaymath}
W_{3} = \left(\begin{array}{cc} 1 & 0 \\ 0 & -1 \end{array}\right), \ \ \ \ W_{4} = \left(\begin{array}{cc} i & 0 \\ 0 & i \end{array}\right).
\end{displaymath}

And the $\bar{W}^{I}_{a\bar{b}}$ are the Hermitian conjugates of these.  In addition, the covariant derivative is defined as
\begin{equation}
D_{\mu} = \partial_{\mu} + ig A_{\mu}
\end{equation}
and the field strength as
\begin{equation}
F_{\mu\nu} = D_{\mu}A_{\nu} - D_{\nu}A_{\mu}.
\end{equation}

Spinor notation follows the conventions of Wess and Bagger.

\end{document}